\documentclass[twocolumn,showpacs,amssymb,prd,superscriptaddress,nofootinbib]{revtex4-1}

\usepackage{graphicx,epsf, epsfig, amssymb}% Include figure files
\usepackage{bm}
\usepackage{longtable}
\usepackage{color}
\usepackage{hyperref}
\usepackage{amsfonts,amsmath,amssymb,mathrsfs}

\def\be{\begin{equation}}
\def\ee{\end{equation}}
\def\beq{\begin{eqnarray}}
\def\eeq{\end{eqnarray}}
\newcommand{\nn}{\nonumber}

\newcommand{\lp}{\left(}
\newcommand{\rp}{\right)}
\newcommand{\lb}{\left[}
\newcommand{\rb}{\right]}

\newcommand{\bgaln}{\begin{align}}
\newcommand{\bgeq}{\begin{equation}}

\newcommand{\Rmnum}[1]{\expandafter\@slowromancap\romannumeral #1@}

\begin{document}

\title{Up to eleven: radiation from particles with arbitrary energy\\falling
  into higher-dimensional black holes}


\author{Emanuele Berti\footnote{Electronic address: berti@phy.olemiss.edu}}

\affiliation{Department of Physics and Astronomy, The University of
  Mississippi, University, MS 38677, USA}

\affiliation{California Institute of Technology, Pasadena, CA 91109, USA}

\author{Vitor Cardoso\footnote{Electronic address: vitor.cardoso@ist.utl.pt}}

\affiliation{CENTRA, Departamento de F\'{\i}sica, Instituto Superior
  T\'ecnico, Universidade T\'ecnica de Lisboa - UTL, Av.~Rovisco Pais 1, 1049
  Lisboa, Portugal}

\affiliation{Department of Physics and Astronomy, The University of
  Mississippi, University, MS 38677, USA}

\author{Barnabas Kipapa\footnote{Electronic address: brkipapa@olemiss.edu}}

\affiliation{Department of Physics and Astronomy, The University of
  Mississippi, University, MS 38677, USA}

\begin{abstract}
We consider point particles with arbitrary energy per unit mass $E$ that fall
radially into a higher-dimensional, nonrotating, asymptotically flat black
hole. We compute the energy and linear momentum radiated in this process as
functions of $E$ and of the spacetime dimensionality $D=n+2$ for $n=2,\dots,9$
(in some cases we go up to $11$). We find that the total energy radiated
increases with $n$ for particles falling from rest ($E=1$). For fixed particle
energies $1<E\leq 2$ we show explicitly that the radiation has a local minimum
at some critical value of $n$, and then it increases with $n$. We conjecture
that such a minimum exists also for higher particle energies. The present
point-particle calculation breaks down when $n=11$, because then the radiated
energy becomes larger than the particle mass. Quite interestingly, for $n=11$
the radiated energy predicted by our calculation would also violate Hawking's
area bound. This hints at a qualitative change in gravitational radiation
emission for $n\gtrsim 11$. Our results are in very good agreement with
numerical simulations of low-energy, unequal-mass black hole collisions in
$D=5$ (that will be reported elsewhere) and they are a useful benchmark for
future nonlinear evolutions of the higher-dimensional Einstein equations.
\end{abstract} 

\maketitle

%%%%%%%%%%%%%%%%%%%%%%%%%%%%%%%%%%%%%%%%%%%%%%%%%%%%%%%%%%%%%%%%%%%%%%%%%%%%%%%
\section{Introduction}
%%%%%%%%%%%%%%%%%%%%%%%%%%%%%%%%%%%%%%%%%%%%%%%%%%%%%%%%%%%%%%%%%%%%%%%%%%%%%%%
The dynamics of black holes (BHs) in generic spacetimes has attracted
considerable attention in recent years. In astrophysics, BHs are important as
sources of gravitational and electromagnetic waves.  The inspiral and merger
of BH binaries is a primary target for Earth-based and space-based
gravitational-wave detectors \cite{Sathyaprakash:2009xs}. In gas-rich
environments, BH mergers may be associated with detectable electromagnetic
precursors or afterglows \cite{Dotti:2006zn,O'Neill:2008dg} and even drive the
production of jets \cite{Palenzuela:2010nf}.  In high-energy physics, the
gauge/gravity duality \cite{Maldacena:1997re,Witten:1998qj} has created a
powerful framework for the study of strongly coupled gauge theories, with
applications in connection with the experimental program on heavy ion
collisions at RHIC \cite{Mateos:2007ay} and LHC \cite{ArkaniHamed:1998rs,Randall:1999vf}, among many others. The fact that BHs
on the gravitational side of this correspondence are dual to thermal states of
the gauge theory has sparked a renewed interest in BH physics. Furthermore,
some proposals to solve the hierarchy problem postulate the existence of extra
dimensions accessible only to gravity
\cite{ArkaniHamed:1998rs,Randall:1999vf}. In these scenarios, BH production
from the collision of particles at energy scales above TeV is an almost
inescapable consequence.

Gravitational wave detection and high-energy applications require an accurate
knowledge of BH dynamics and gravitational radiation emission.  This triggered
research on the numerical evolution of the full nonlinear Einstein equations
in four
\cite{Pretorius:2005gq,Buonanno:2006ui,Berti:2007fi,Aylott:2009ya,Sperhake:2008ga,Sperhake:2009jz,Campanelli:2007ea,Ponce:2010fq,Witek:2010qc}
and higher dimensions
\cite{Yoshino:2009xp,Zilhao:2010sr,Witek:2010xi,Shibata:2010wz,Sorkin:2009wh}. The
validation of numerical codes requires semianalytical tools, such as
post-Newtonian theory, BH perturbation theory and zero-frequency expansions to
model BH collisions.  Such tools have been available for decades in the case
of four-dimensional, asymptotically flat spacetimes (see
e.g. \cite{Berti:2010ce} and references therein). The same cannot be said of
$D$-dimensional spacetimes, but recently there has been significant progress
in this field. For instance, Refs.~\cite{Cardoso:2002pa,Cardoso:2008gn}
investigated gravitational radiation and the quadrupole formalism in
higher-dimensional, asymptotically flat spacetimes. These studies showed that
odd- and even-dimensional spacetimes behave differently, but there are simple
energy formulas in the Fourier-domain that apply to both cases
\cite{Cardoso:2008gn}.

Linearized perturbations of higher-dimensional BHs are now well understood
\cite{Kodama:2003kk,Kodama:2003jz,Ishibashi:2003ap,Durkee:2010qu}.
Historically, perturbative methods such as the close-limit approximation
\cite{Price:1994pm} (recently extended to higher dimensions
\cite{Yoshino:2005ps,Yoshino:2006kc}) have provided guidance and insight in
the numerical analysis of BH mergers in general relativity.  The application
of higher-dimensional BH perturbation theory to compute gravitational
radiation in situations of physical interest was initiated in
Ref.~\cite{Berti:2003si} (henceforth Paper I), where the authors studied the
radiation produced by ultrarelativistic particles falling into
even-dimensional, nonrotating, asymptotically flat BHs.

\begin{figure*}[htb]
\begin{center}
\begin{tabular}{cc}
\includegraphics[scale=0.33,clip=true]{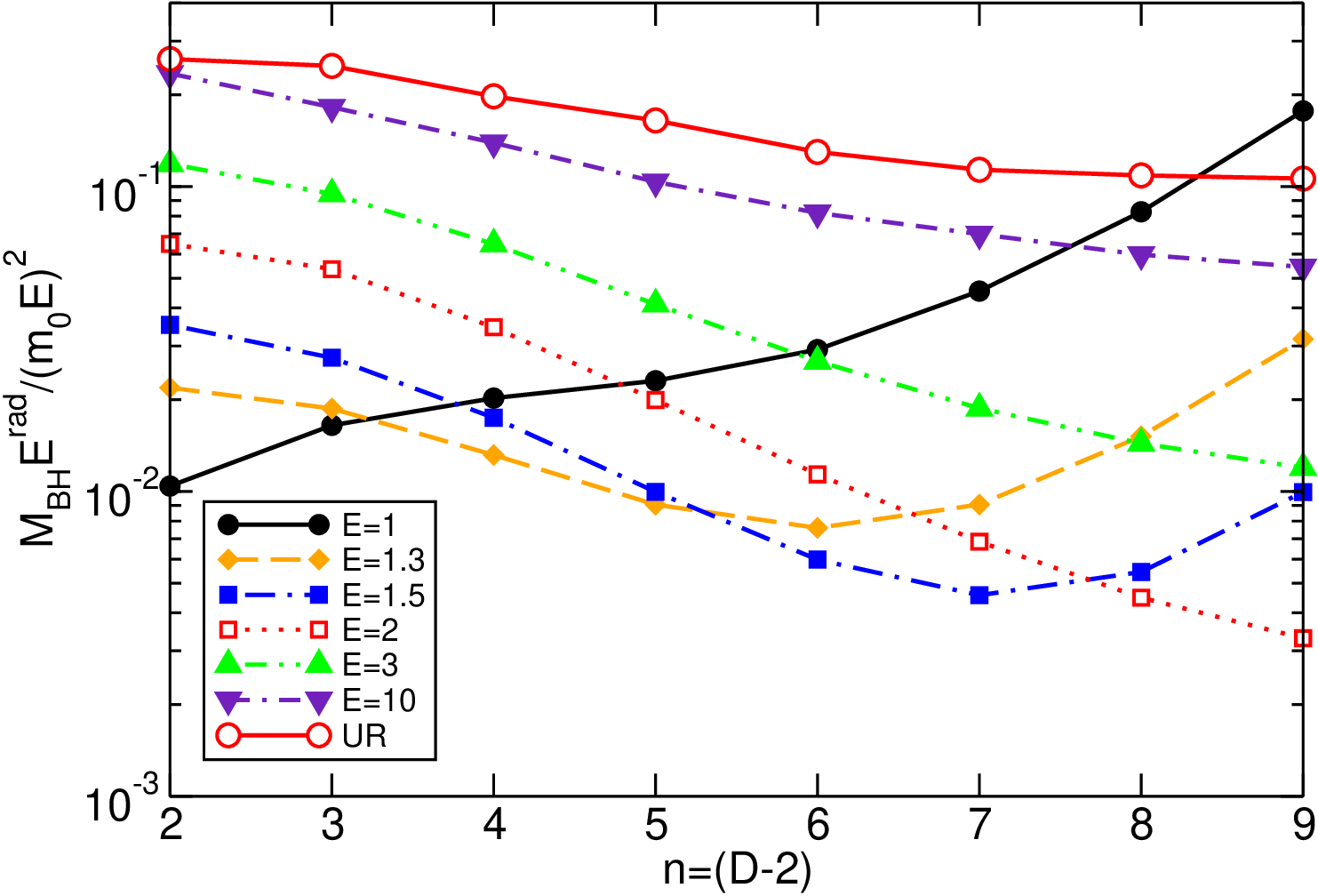}
&\includegraphics[scale=0.33,clip=true]{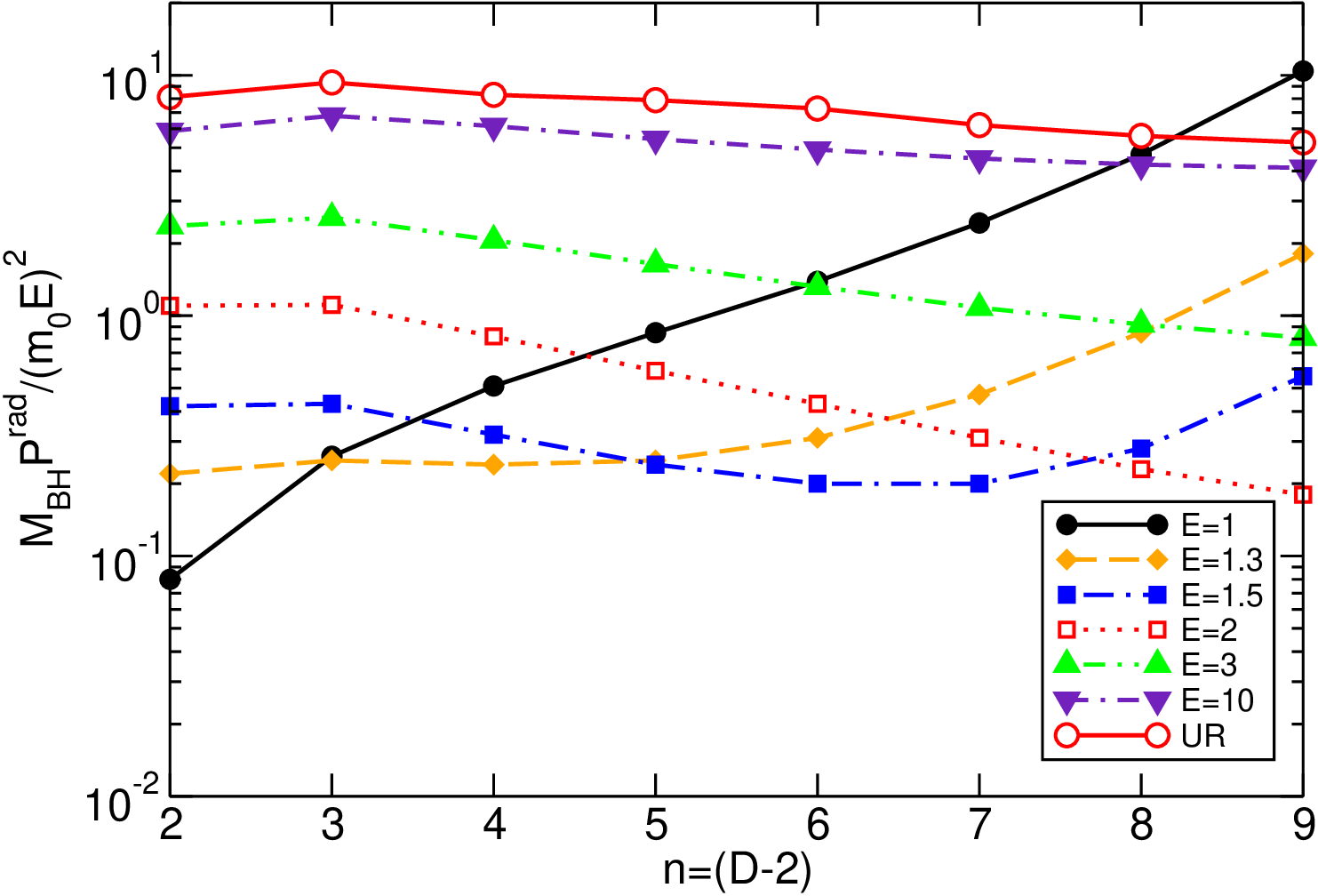}\\
\end{tabular}
\caption{\label{fig:energymom} Energy and momentum radiated plotted as
  functions of $n$ for selected values of $E$.}
\end{center}
\end{figure*}

Numerical codes to evolve the Einstein equations in higher dimensions are
presently capable of handling low-energy BH collisions in five dimensions
\cite{Witek:2010xi}. However, the extension of these results to high-energy
collisions in spacetimes of generic dimensionality presents a significant
challenge.  Motivated by these developments, here we extend the analysis of
Paper I to study the energy and linear momentum radiated when particles of
{\em arbitrary energy} fall into nonrotating, higher-dimensional BHs. Our
results are in remarkable agreement with five-dimensional simulations of
unequal-mass BH collisions in higher dimensions, that will be reported
elsewhere \cite{witek2010}. They also provide useful (and sometimes
surprising) insight into the energy- and dimensionality-dependence of
gravitational radiation produced by head-on BH collisions.

The main findings of this paper are summarized in Figure \ref{fig:energymom},
where we show the radiated energy (left) and linear momentum (right) as a
function of $n=D-2$ for selected values of the particle energy per unit mass
$E$ ($E=1,\,1.3,\,1.5,\,2,\,3,\,10$). Infalls from rest correspond to $E=1$,
and the ultrarelativistic case $E\to \infty$ is denoted by ``UR'' in the
legend. As natural in perturbation theory, the energy and angular momentum
radiated are inversely proportional to the BH mass $M_{\rm BH}$ and
proportional to the square of the particle energy in the UR limit, so in the
plot we normalize the radiation to $(m_0E)^2/M_{\rm BH}$. Paper I found that
the radiated energy {\em decreases} with $n$ for ultrarelativistic infalls
with $n\leq 8$\footnote{An apparent exception to this rule is the case $n=8$
  in Table VI of Paper I. Unfortunately, the extrapolated energy for $n=8$
  ($D=10$) was overestimated by $\sim 20\%$ in that paper. The reason is that
  we ``only'' computed multipoles up to $l=20$ to estimate the total
  radiation, and as it turns out, this was not enough to get a reliable
  extrapolation of the total radiated energy. This error has been fixed here
  (see Table \ref{tab:energy} below).}. Figure \ref{fig:energymom} shows that
the total energy radiated {\em increases} with $n$ for particles falling from
rest. Our results for $E=1$ and $n=3$ are in remarkably good agreement with
numerical simulations of low-energy, unequal-mass BH collisions in $D=5$
\cite{witek2010}. They should also provide a useful benchmark for future
nonlinear evolutions of the Einstein equations in higher dimensions.

Even more interestingly, in some cases the left panel of Figure
\ref{fig:energymom} shows the existence of a {\em local minimum} of the
radiation as a function of $n$. This minimum is visible in the plot for the
cases when the infall is not kinetic-energy dominated ($E=1.3$ and $E=1.5$),
but we verified that it also occurs for $E=2$ by extending our calculation to
$n=11$. {\em We conjecture that such a local minimum exists for {\em any}
  $E>1$, and that the radiated energy may generically increase for
  sufficiently large $n$, eventually violating the point-particle
  approximation and the area theorem bound.} Past work showed that
point-particle results in four dimensions can be successfully extrapolated to
the comparable-mass case (see e.g.~\cite{Anninos:1995vf}).  Our results imply
that for large $n$ this will no longer be the case.  In fact, we find that
when $n=11$ the radiation emitted in infalls from rest is {\em larger} than
the particle mass, and therefore the point-particle approximation must break
down. Incidentally, this breakdown fits in nicely with Hawking's area bound
\cite{Hawking:1971tu}.
Hawking's area theorem, applied to infalls from rest in generic spacetime
dimension $n$, predicts that the amount of radiation emitted in equal-mass BH
collisions decreases with $n$. Our results do not violate the area
theorem. Instead, they suggest a failure of the point-particle approach for a
change in behavior of the total radiated energy (in the equal-mass case) for
$n\gtrsim 11$. Unlike {\em Spinal Tap}'s Nigel Tufnel, in higher-dimensional
gravitational radiation we can never ``go to eleven''.

The plan of the paper is as follows. In Section \ref{eqs} we briefly recall
how to compute the radiation produced by particles falling radially (but with
arbitrary energy) into a higher-dimensional BH. In Section \ref{results} we
present our results on the radiated energy and linear momentum (``kick
velocity''), along with a preliminary comparison with numerical relativity
results that will be presented in a companion paper. Section \ref{concl}
contains conclusions and possible directions for future research. In Appendix
\ref{app:source} we collect, for reference, some technical results.

\begin{figure}[thb]
\includegraphics[scale=0.33,clip=true]{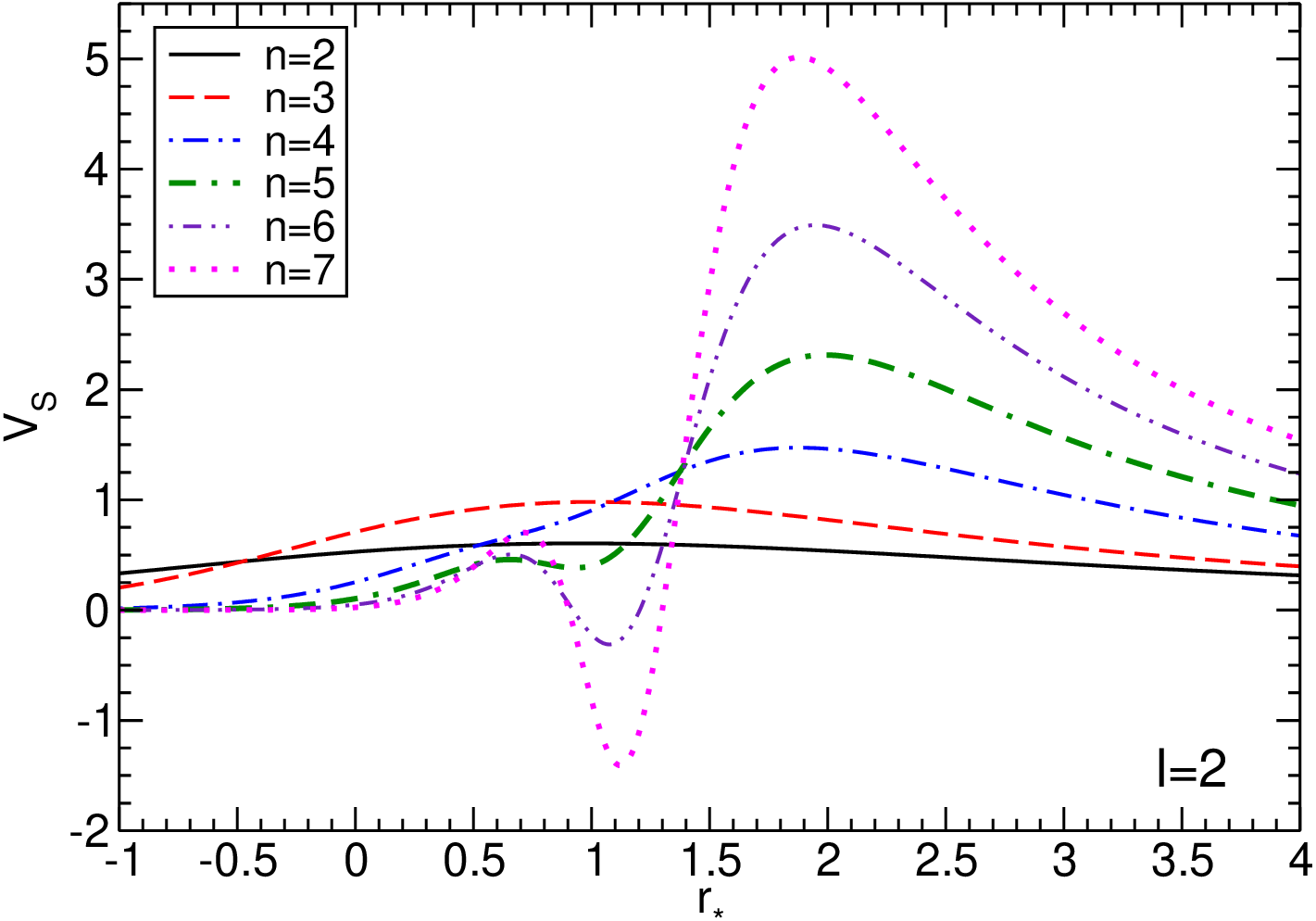}
\caption{\label{Vkodama} The scalar potential of Eq.~(\ref{pscalar}) as a
  function of the tortoise coordinate $r_*$ for $l=2$ and selected values of
  $n$. We use units $r_h=1$.}
\end{figure}
%

%%%%%%%%%%%%%%%%%%%%%%%%%%%%%%%%%%%%%%%%%%%%%%%%%%%
\section{\label{eqs}Formulation of the problem}
%%%%%%%%%%%%%%%%%%%%%%%%%%%%%%%%%%%%%%%%%%%%%%%%%%%

The spherically symmetric BH in $D=n+2$ dimensions is described by the
Schwarzschild-Tangherlini metric \cite{Tangherlini:1963bw}
\be\label{schw}
ds^2=-f(r)dt^2+\frac{dr^2}{f(r)}+r^2 d\Omega_n^2\,,
\ee
where $d\Omega_n$ is the metric of the $n$-dimensional unit sphere $S^n$, and 
\be\label{fdef}
f(r)=1-\frac{2M}{r^{n-1}}\,.
\ee
The BH mass is related to the parameter $M$ by
\be\label{mass}
M_{\rm BH}=\frac{nM{\cal A}_n}{8\pi c^2 G_{n+2}}\,,
\ee
where ${\cal A}_n=2\pi^{(n+1)/2}/\Gamma[(n+1)/2]$ is the area of $S^n$,
$G_{n+2}$ is the $(n+2)$-dimensional Newton constant, and $c$ is the speed of
light. We will set $G_{n+2}=1$ and $c=1$ in the following. The tortoise
coordinate $r_*$ is defined by
\be\label{tort1}
\frac{dr_*}{dr}=\frac{1}{f(r)}\,.
\ee
An analytical expression for $r_*(r)$ valid for generic $n$ is given in Paper
I, Eqs.~(5) and (6). Here and throughout the paper we use the notation of
Ref.~\cite{Kodama:2003jz}.

The computation of the gravitational wave emission of an ultrarelativistic
particle plunging into a BH requires the numerical integration of the
inhomogeneous wave equation for scalar gravitational perturbations (``vector''
and ``tensor'' gravitational perturbations, in the terminology of Kodama and
Ishibashi, are not excited by a particle in radial infall). Setting $x\equiv
2M/r^{n-1}$, the equation for the scalar perturbations is
\be\label{inhom}
\left(\frac{d^2}{dr_*^2}+\omega^2-V_S\right)\Phi_l^{(n)}=S_l^{(n)}\,.  
\ee
where the scalar potential $V_S$ is plotted in Figure \ref{Vkodama} for
selected values of $n$ and $l=2$. This potential is given by
\be\label{pscalar}
V_S=\frac{f(r)Q(r)}{16r^2H(r)^2}\,,
\ee
where the function
\be
H(r)=m+\frac{n(n+1)x}{2}
\ee
with $x=2M/r^{n-1}$,
$m=\kappa^2-n$,
$\kappa^2=l(l+n-1)$,
and
\beq
Q(r)&=&n^4(n+1)^2x^3+n(n+1)\times \nn\\
&\times&[4(2n^2-3n+4)m+n(n-2)(n-4)(n+1)]x^2\nn\\
&-&12n[(n-4)m+n(n+1)(n-2)]mx\nn\\
&+&16m^3+4n(n+2)m^2\,.
\eeq

To simplify the notation, below we will omit the superscript $(n)$ from the
wavefunction $\Phi_l^{(n)}$. 

Equation (\ref{inhom}) reduces to the inhomogeneous Zerilli equation
\cite{Zerilli:1971wd} for $n=2$.
The source term $S_l^{(n)}$ in $(n+2)$ dimensions can be calculated from the
stress-energy tensor of the infalling particle. Denote by $E$ the particle
energy per unit mass. Making use of the geodesic equations for massive
particles in radial infall
%
%\begin{align}
%&
\be
\frac{dt}{d\tau}=\frac{E}{f(r)}\,,\quad
%&
\frac{dr}{d\tau}=-\sqrt{E^2-f(r)}\,,
\ee
%\end{align}
%
a straightforward generalization of the calculation presented in Paper I
yields
\begin{widetext}
\be
S_l^{(n)}=\sqrt{32\pi}m_0\mathcal{S}^{nl}e^{i\omega t(r)} \frac{f(r)}{r^{n/2}H} 
\left\{ \frac{E}{i\omega r}\lp
4-\frac{n^2(n+1)[1-f(r)]-2(n-2)m}{H}\rp+\frac{2}{\sqrt{E^2-f(r)}}
\right\}\,.
\label{Gensource}
\ee
\end{widetext}
The normalized Gegenbauer polynomials $\mathcal{S}^{nl}$ are listed for the
relevant values of $n$ in Appendix \ref{app:source}, along with simplified
expressions of the source term in the ultrarelativistic case ($E\to \infty$).

We use a straightforward modification of the {\sc Fortran} code described in
Paper I to solve Eq.~(\ref{inhom}) via Green's function techniques. We refer
the reader to that paper for details. Just like in Paper I, for convenience,
we set the horizon radius $r_h=(2M)^{1/(n-1)}=1$ in our numerical
integrations.  The energy spectrum can be expressed in terms of the wave
amplitude at infinity $\Phi_{l}$, given in Eq.~(20) of Paper I, as
\be\label{energyspec}
\frac{dE_l}{d\omega}=\frac{\omega^2}{16\pi}
\frac{n-1}{n}\kappa^2(\kappa^2-n)|\Phi_{l}|^2\,.
\ee
\begin{figure*}[htb]
\begin{center}
\begin{tabular}{ccc}
\includegraphics[scale=0.23,clip=true]{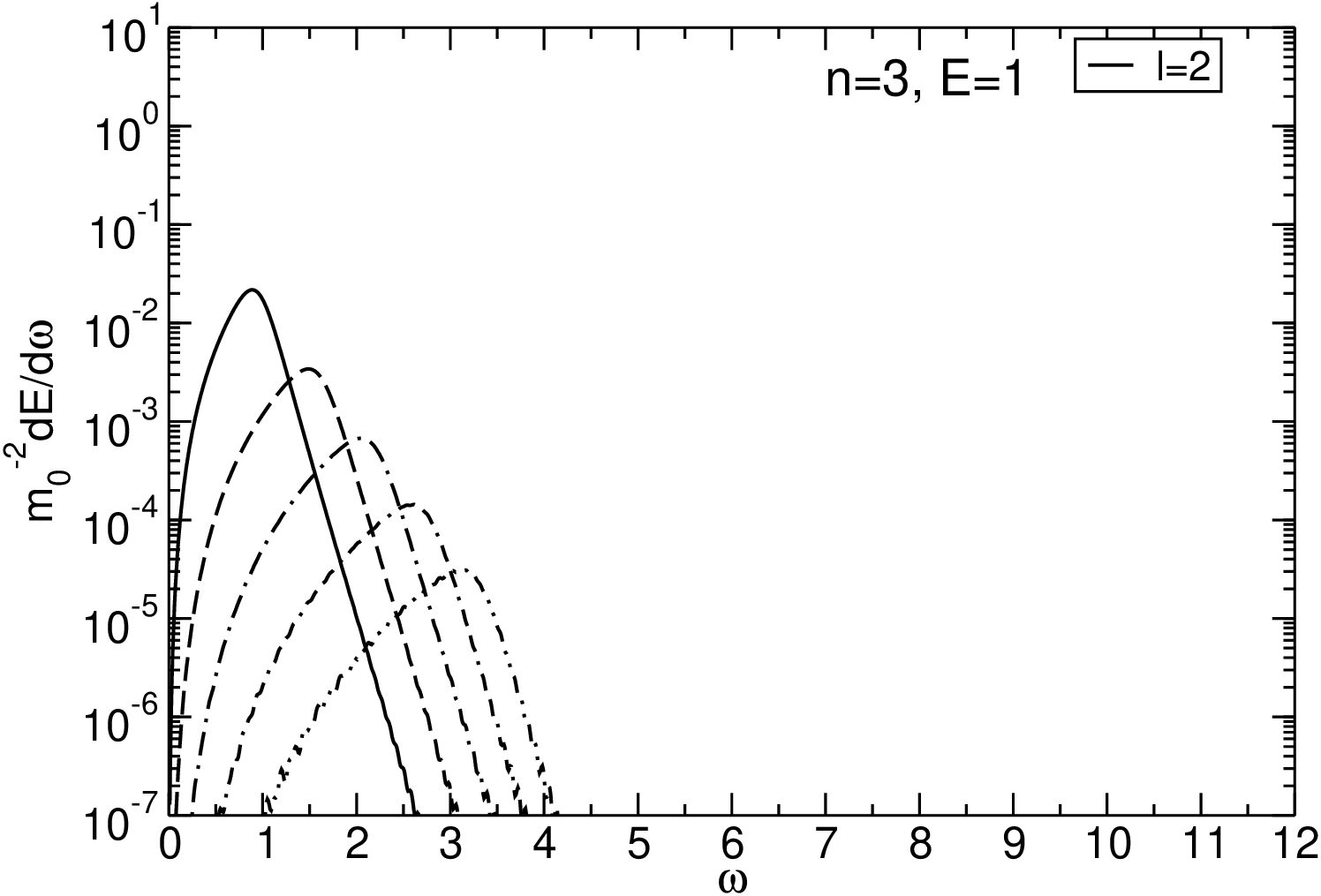}
&\includegraphics[scale=0.23,clip=true]{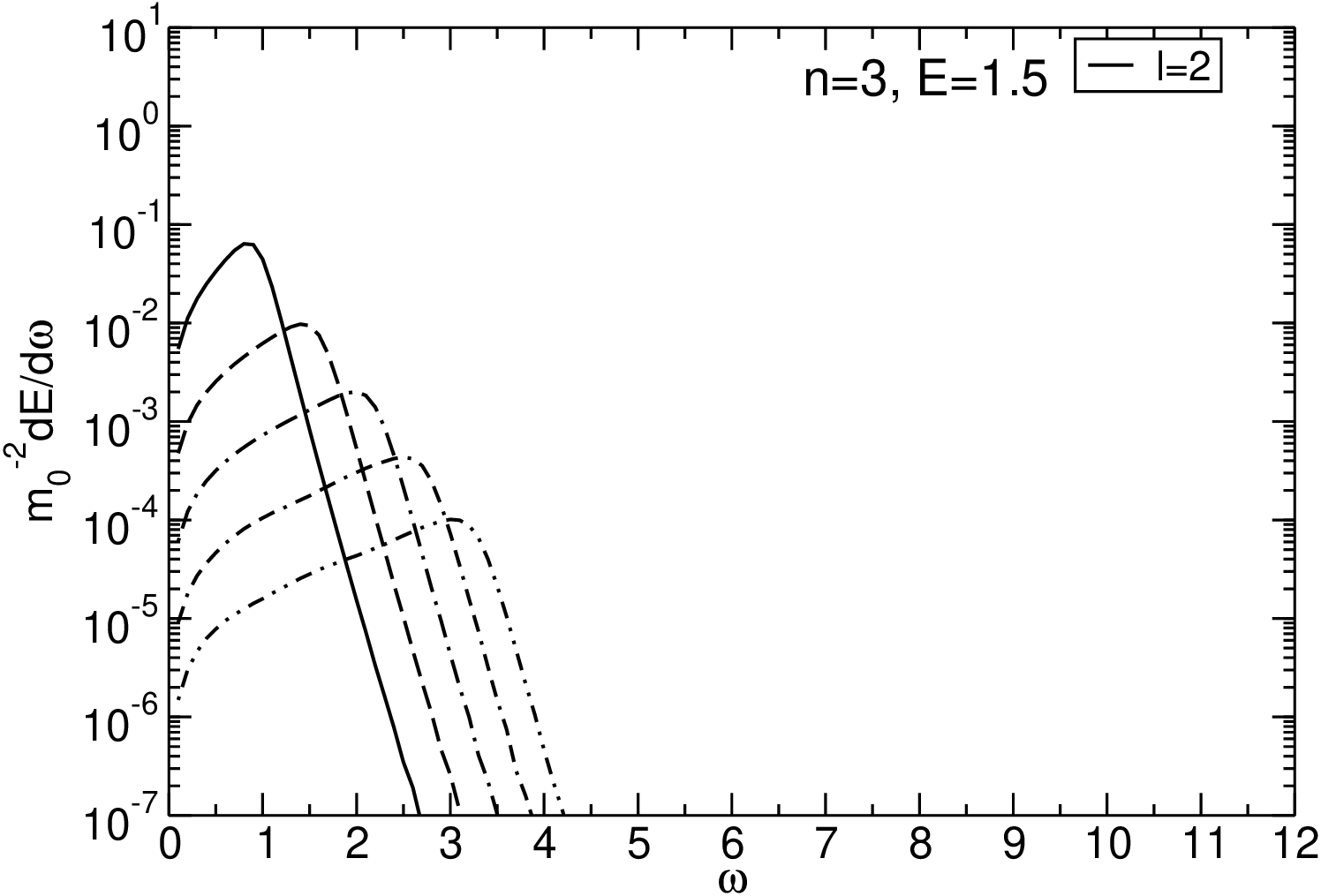} 
&\includegraphics[scale=0.23,clip=true]{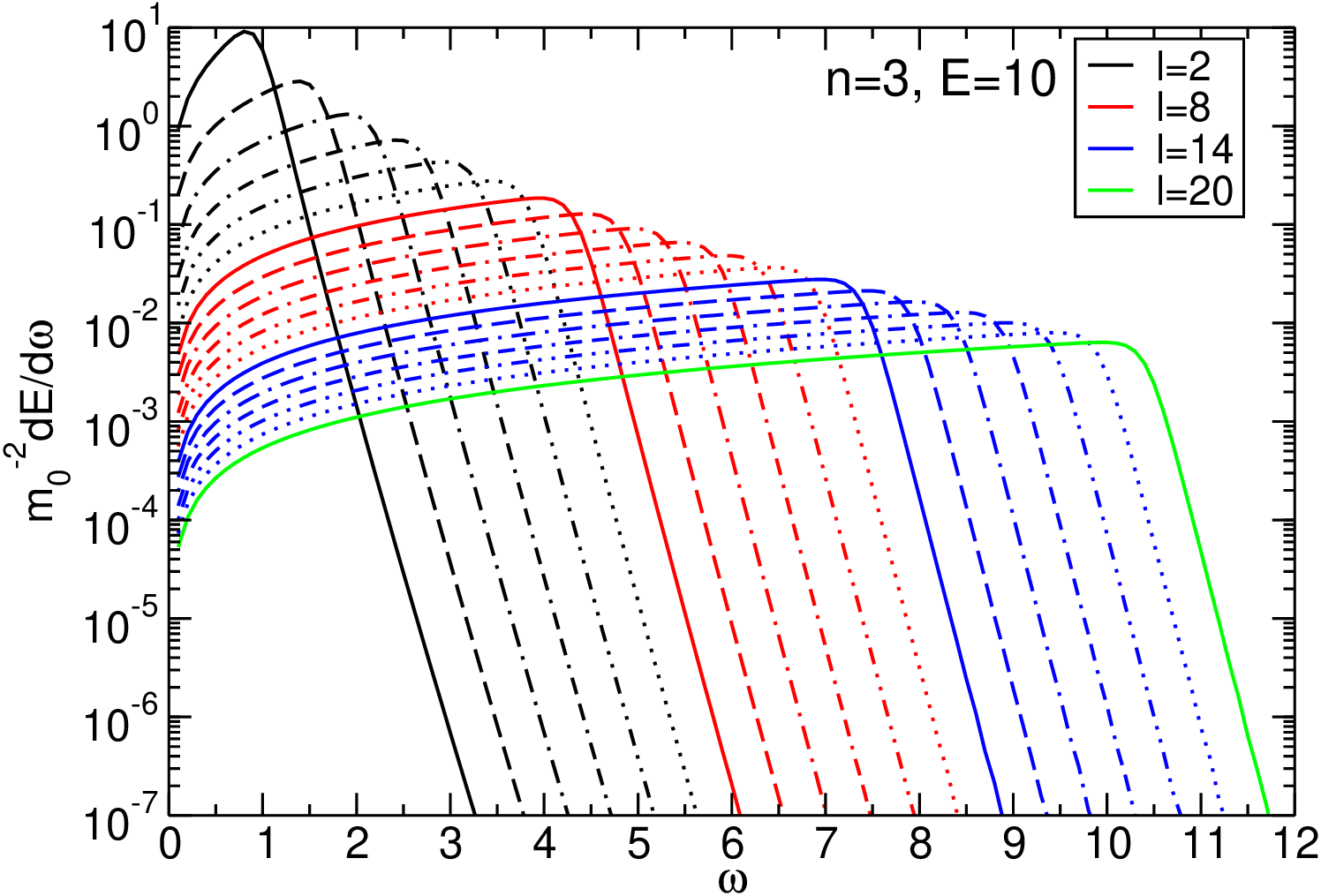} \\
\includegraphics[scale=0.23,clip=true]{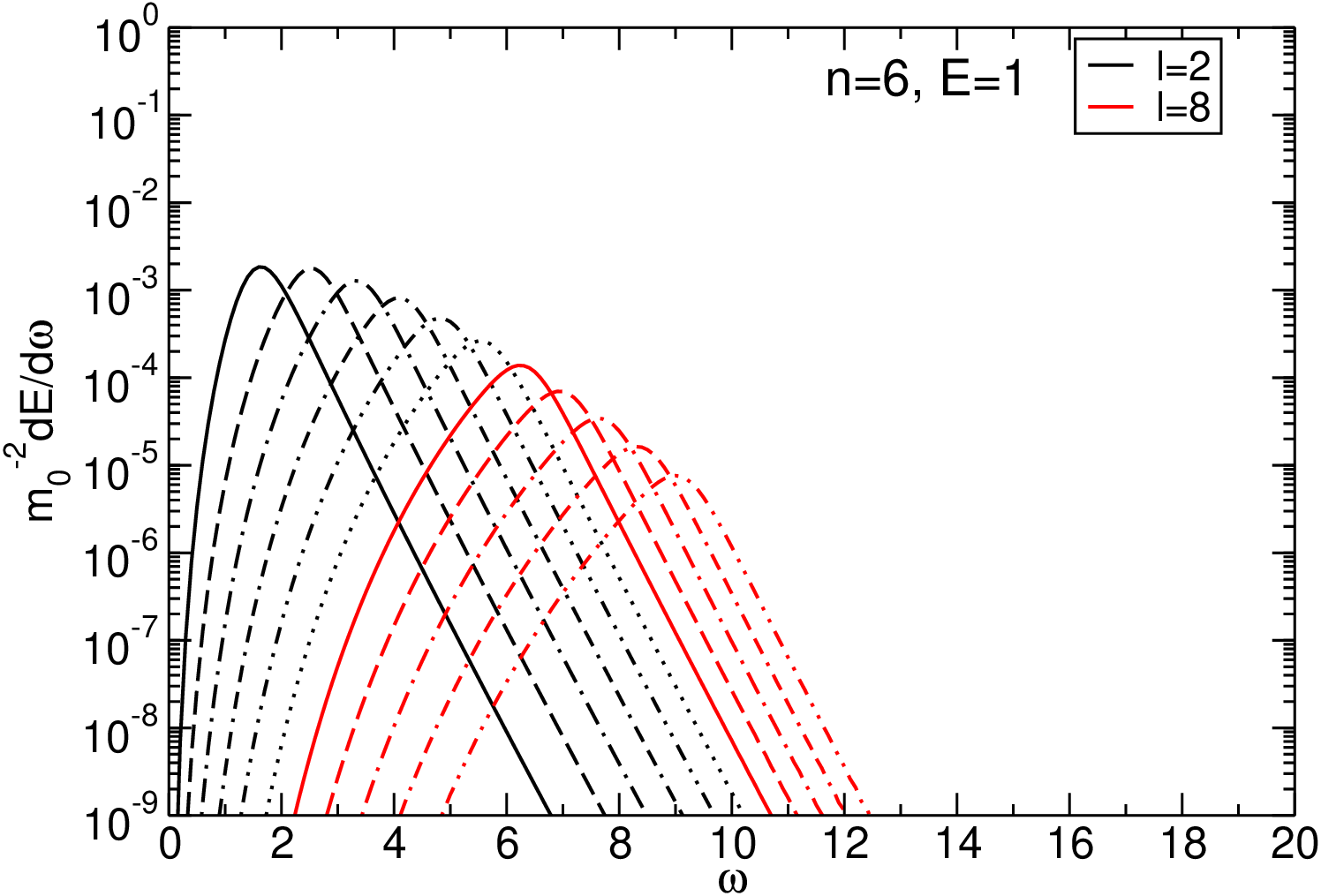}
&\includegraphics[scale=0.23,clip=true]{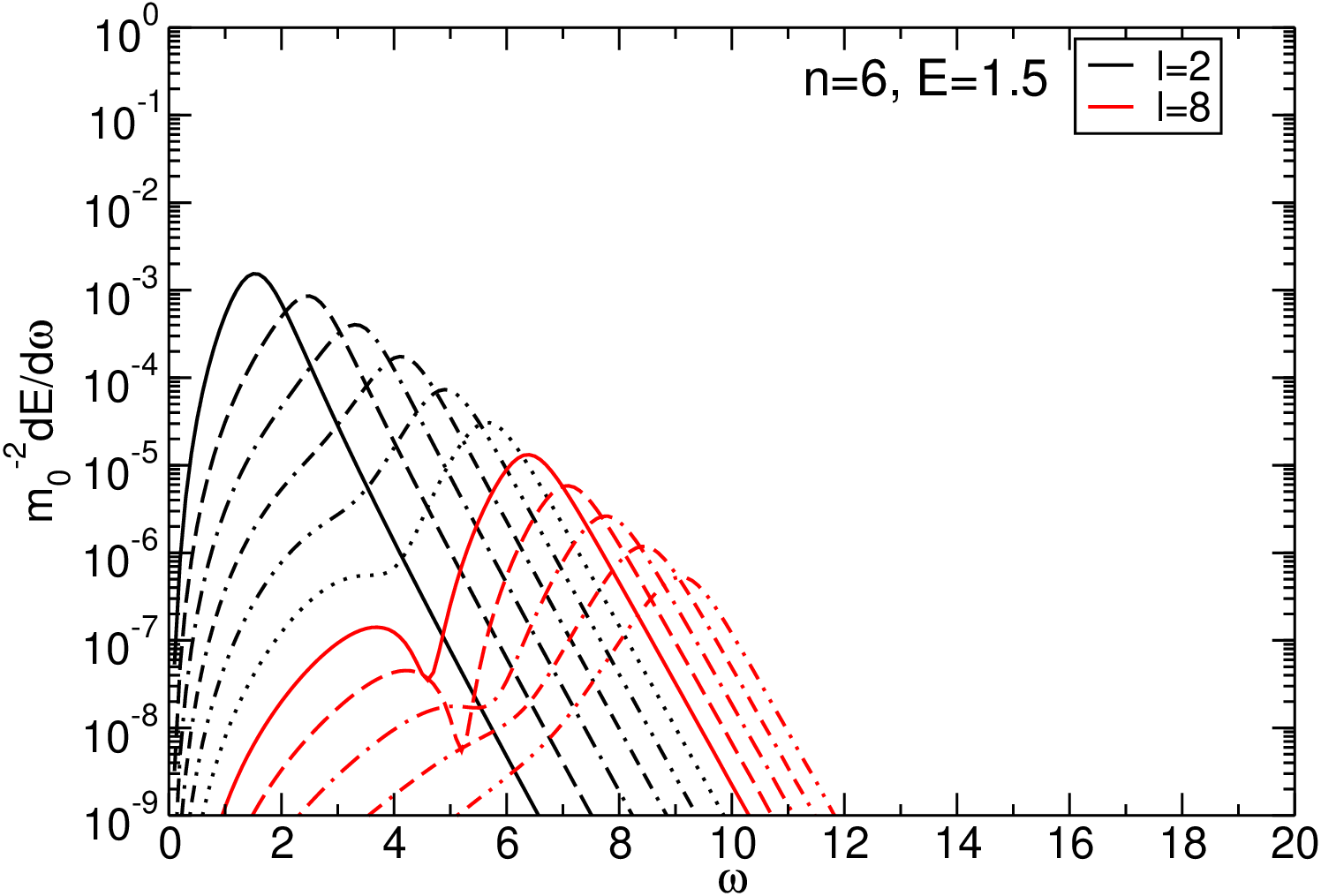}
&\includegraphics[scale=0.23,clip=true]{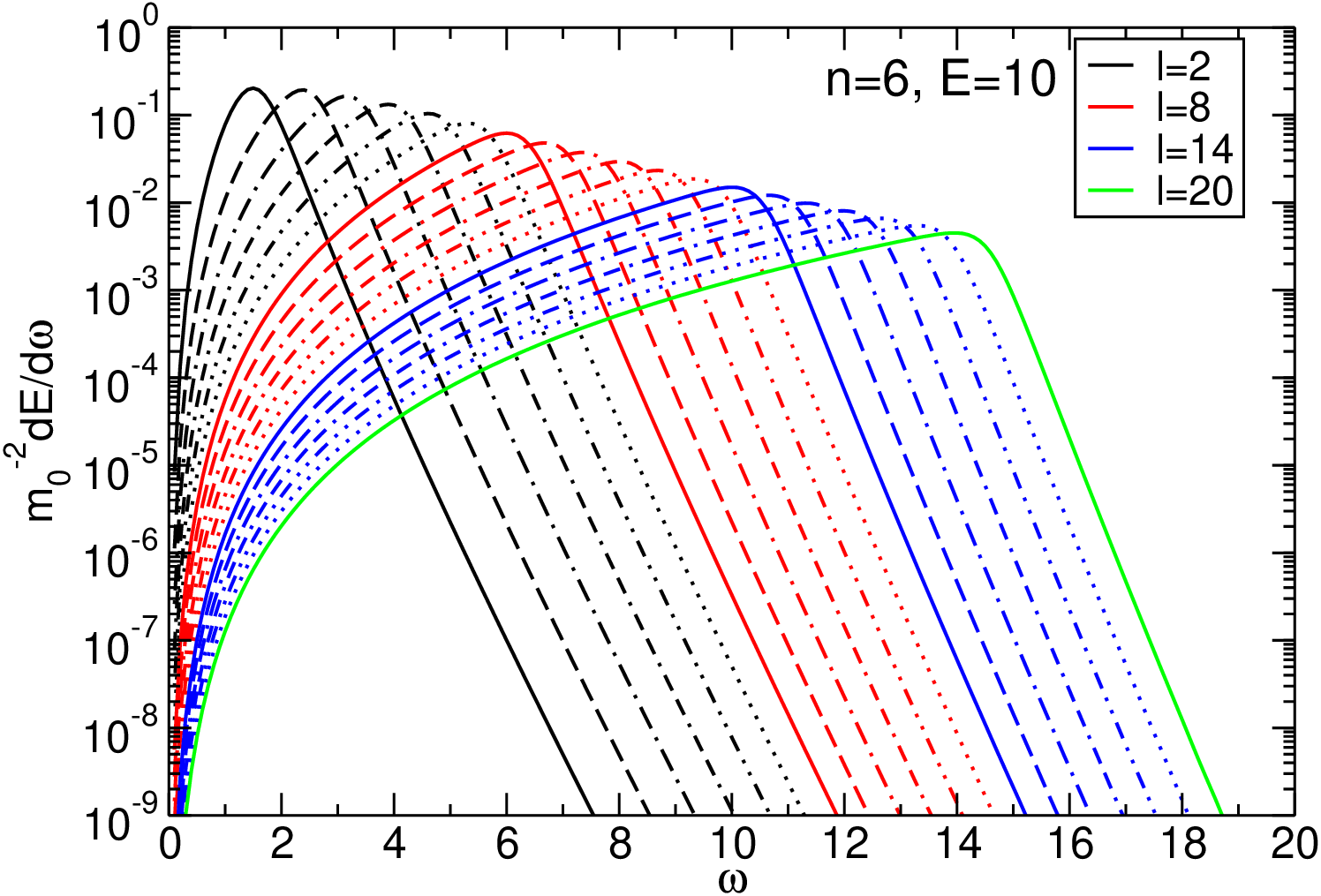} \\
\end{tabular}
\caption{\label{n3spectra} Energy spectra for $n=3$ and $n=6$ (in units
  $r_h=1$).}
\end{center}
\end{figure*}

Paper I did not provide a calculation of the radiated linear momentum
$P^i$. The spectrum of the radiated momentum can be obtained from
\be
\frac{dP^i}{d\omega}=\int_{S_{\infty}}d\Omega \frac{d^2E}{d\omega d\Omega}n^i\,,
\ee
with $n^i$ a unit radial vector on the sphere at infinity $S_{\infty}$.  This
results in an infinite series coupling different multipoles.  Using only the
first two terms in the series, we find for instance
\be
\label{P4}
\frac{dP^z}{d\omega}=3\omega^2
\frac{\sqrt{5}\left(\Phi_{3}\Phi^{*}_{2}+\Phi^*_{3}\Phi_{2}\right)
+10\left(\Phi_{3}\Phi^{*}_{4}+\Phi^*_{3}\Phi_{4}\right)}{4\pi\sqrt{7}}\,
\ee
and
\be
\label{P5}
\frac{dP^z}{d\omega}=\omega^2\frac{5\left(\Phi_{3}\Phi^{*}_{2}
+\Phi^*_{3}\Phi_{2}\right)+21\left(\Phi_3\Phi^{*}_{4}
+\Phi^{*}_3\Phi_{4}\right)}{4\pi}\,
\ee
in $D=4$ and $D=5$, respectively. Here, $\Phi_l$ denotes the $l-$pole
component of the Kodama-Ishibashi wavefunction and an asterisk denotes complex
conjugation. We are assuming one-sided spectra. To get the total radiated
linear momentum $P^{\rm rad}$, in this work we do {\it not} truncate the
series at the order shown in Eqs.~(\ref{P4}) and (\ref{P5}). Instead we sum
the required number of multipoles (typically $\sim 10-20$) to get the desired
accuracy.

\section{\label{results}Results}

Our {\sc Fortran} code passed several code checks. The spectra for $n=2$ are
in excellent agreement with those of Refs.~\cite{Ruffini:1973ky,Berti:2010ce}
for generic energies, and with those of Ref.~\cite{Cardoso:2002ay} in the
ultrarelativistic limit; they have been reported several times in the
literature, so we do not reproduce them here. Our even-dimensional
ultrarelativistic spectra obviously reduce to those shown in Paper I. Results
from the {\sc Fortran} code were also verified by comparison with a {\sc
  Mathematica} notebook.

%%%%%%%%%%%%%%%%%%%%%%%%%%%%%%%%%%%%%%%%%%%%%%%%%%%%%%%%%%%%%%%%%%%%%
\subsection{Energy}
%%%%%%%%%%%%%%%%%%%%%%%%%%%%%%%%%%%%%%%%%%%%%%%%%%%%%%%%%%%%%%%%%%%%%

%
\begin{figure*}[htb]
\begin{center}
\begin{tabular}{ccc}
\includegraphics[scale=0.23,clip=true]{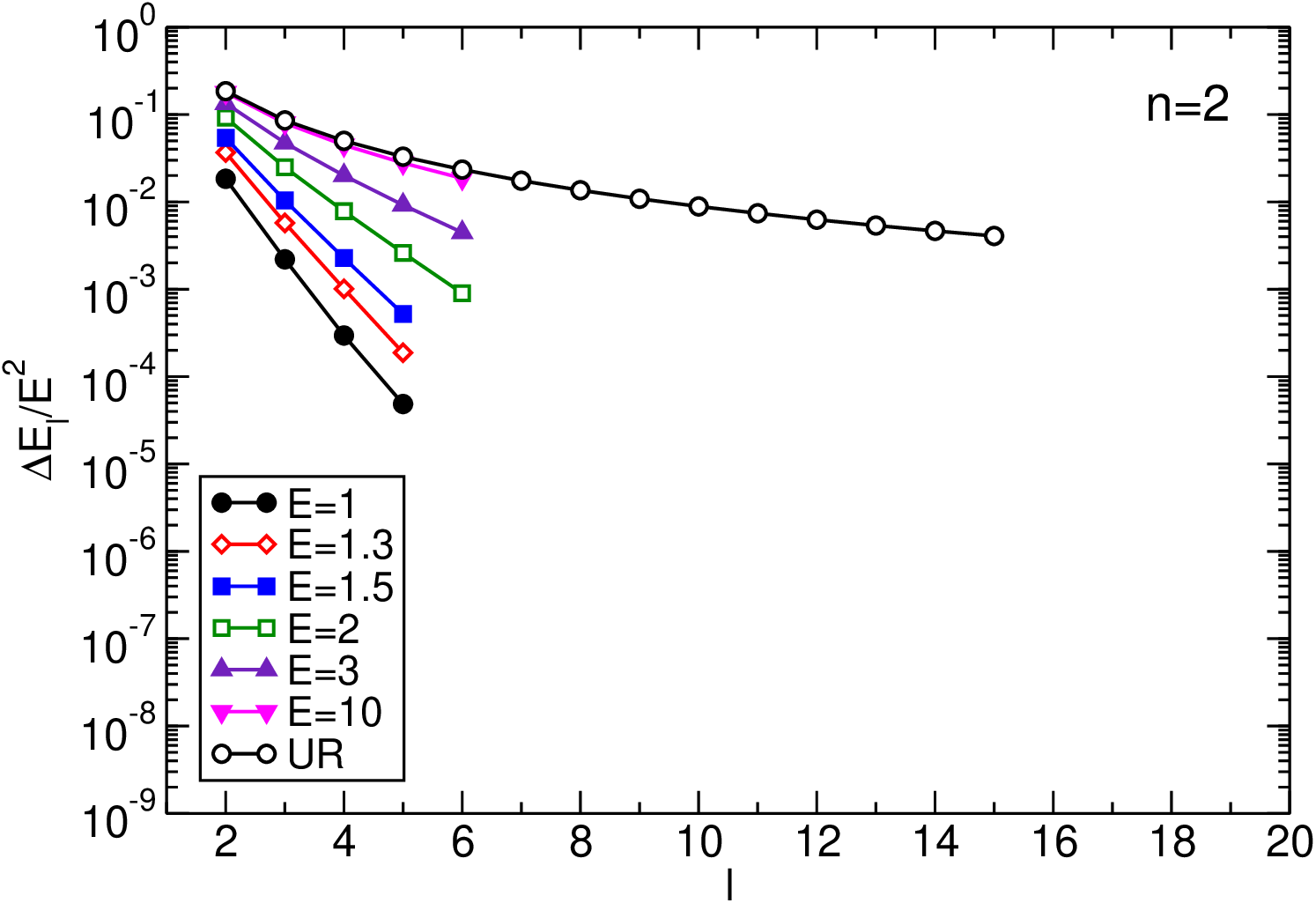}
&\includegraphics[scale=0.23,clip=true]{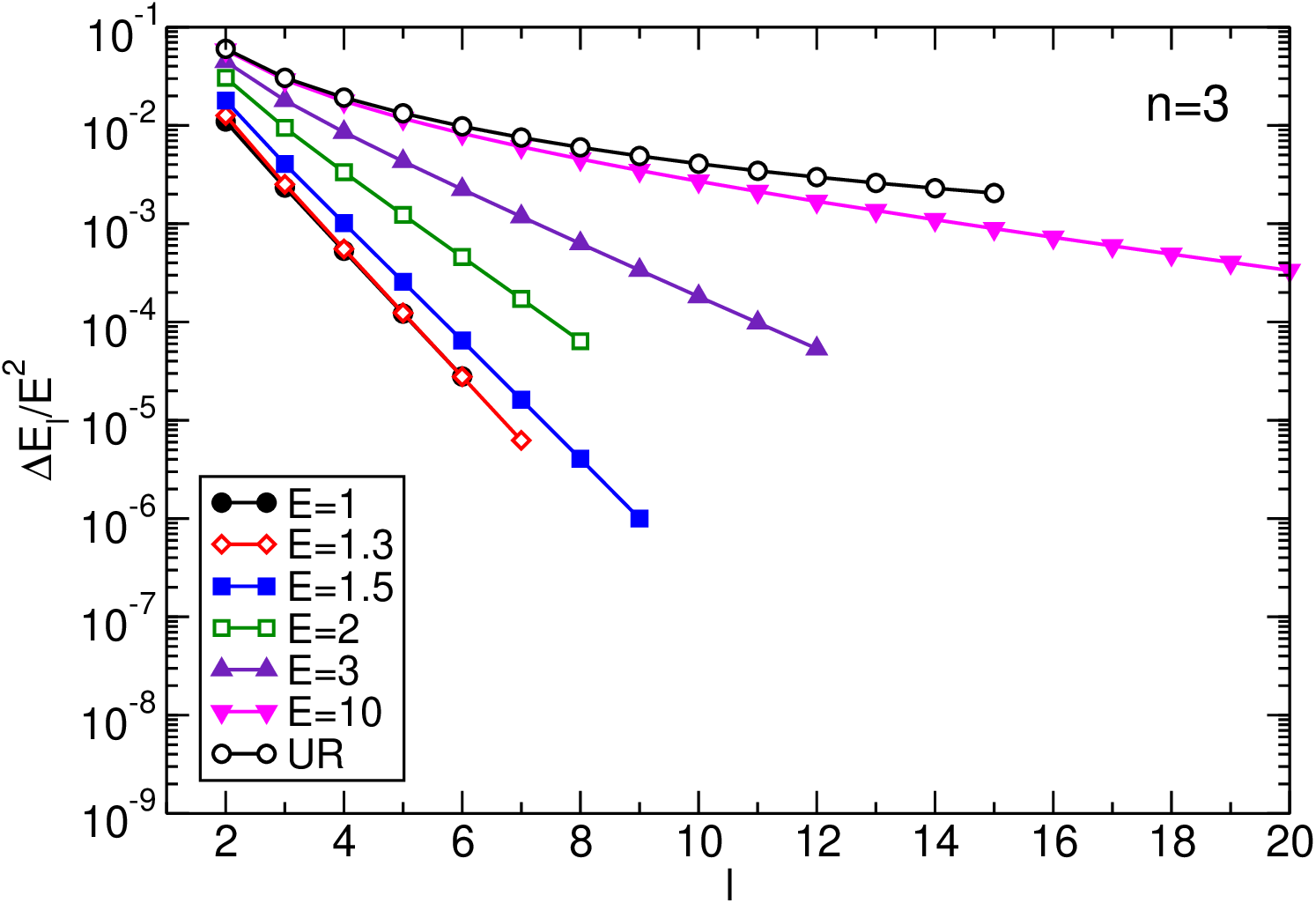}
&\includegraphics[scale=0.23,clip=true]{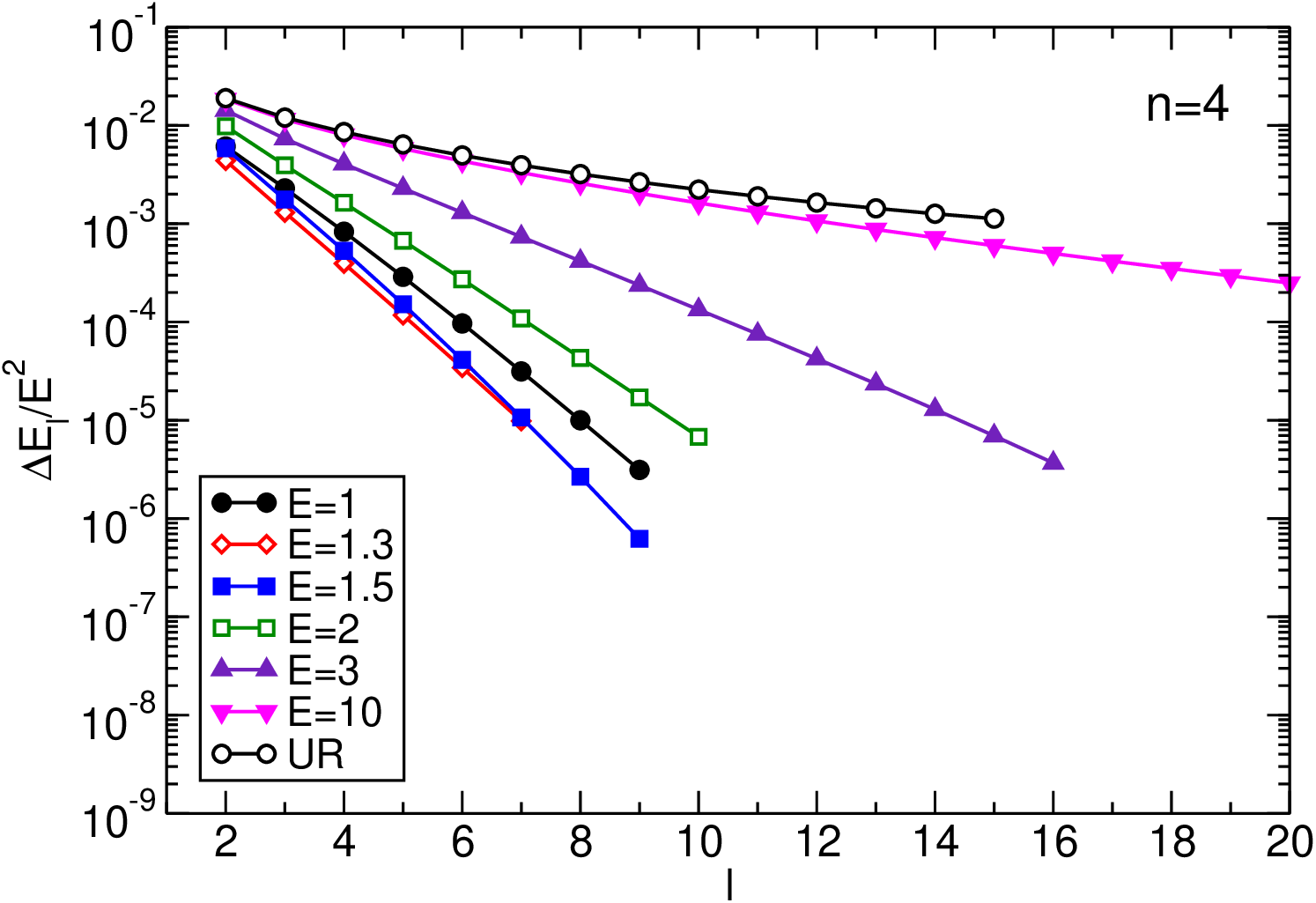}\\
\includegraphics[scale=0.23,clip=true]{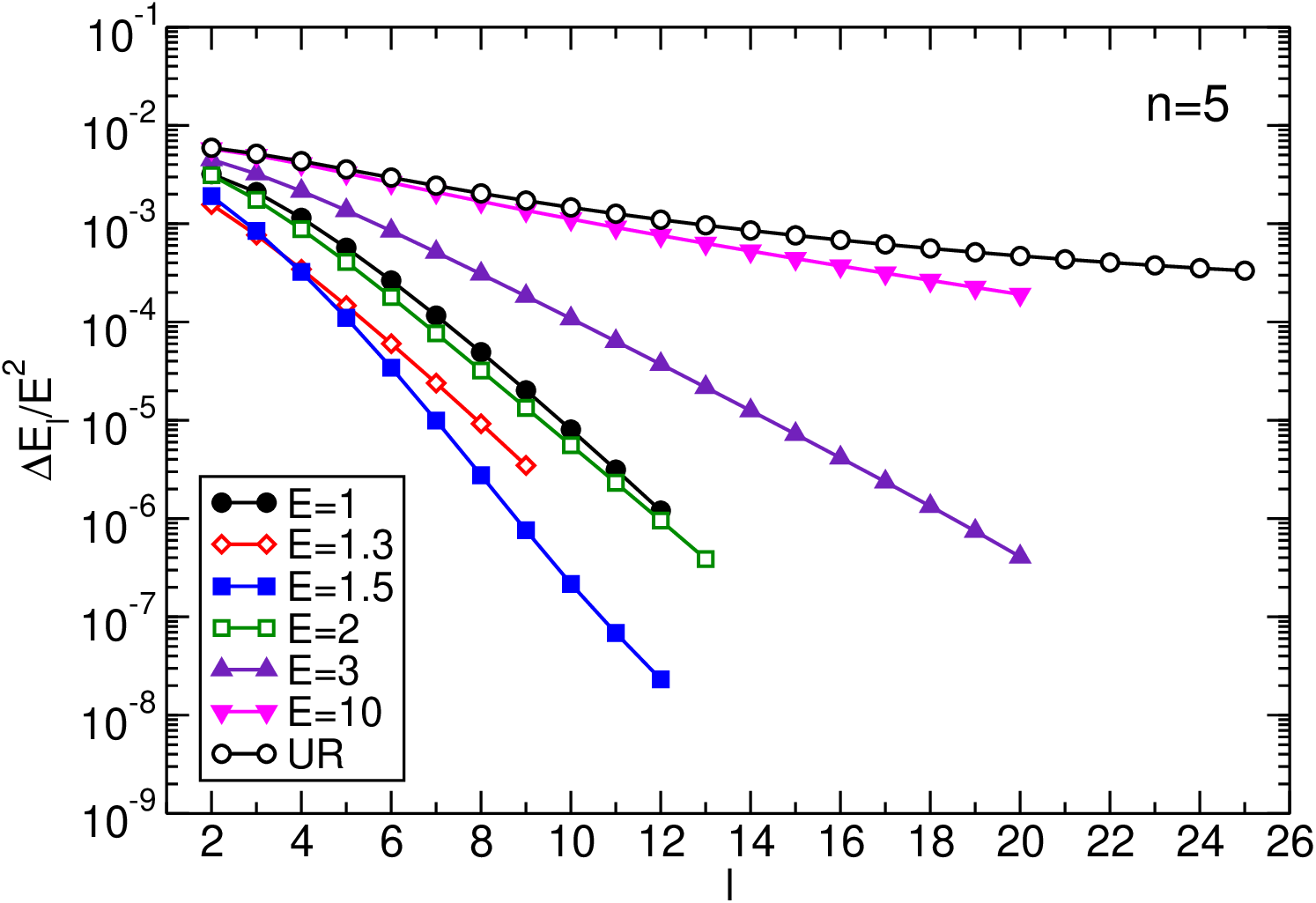}
&\includegraphics[scale=0.23,clip=true]{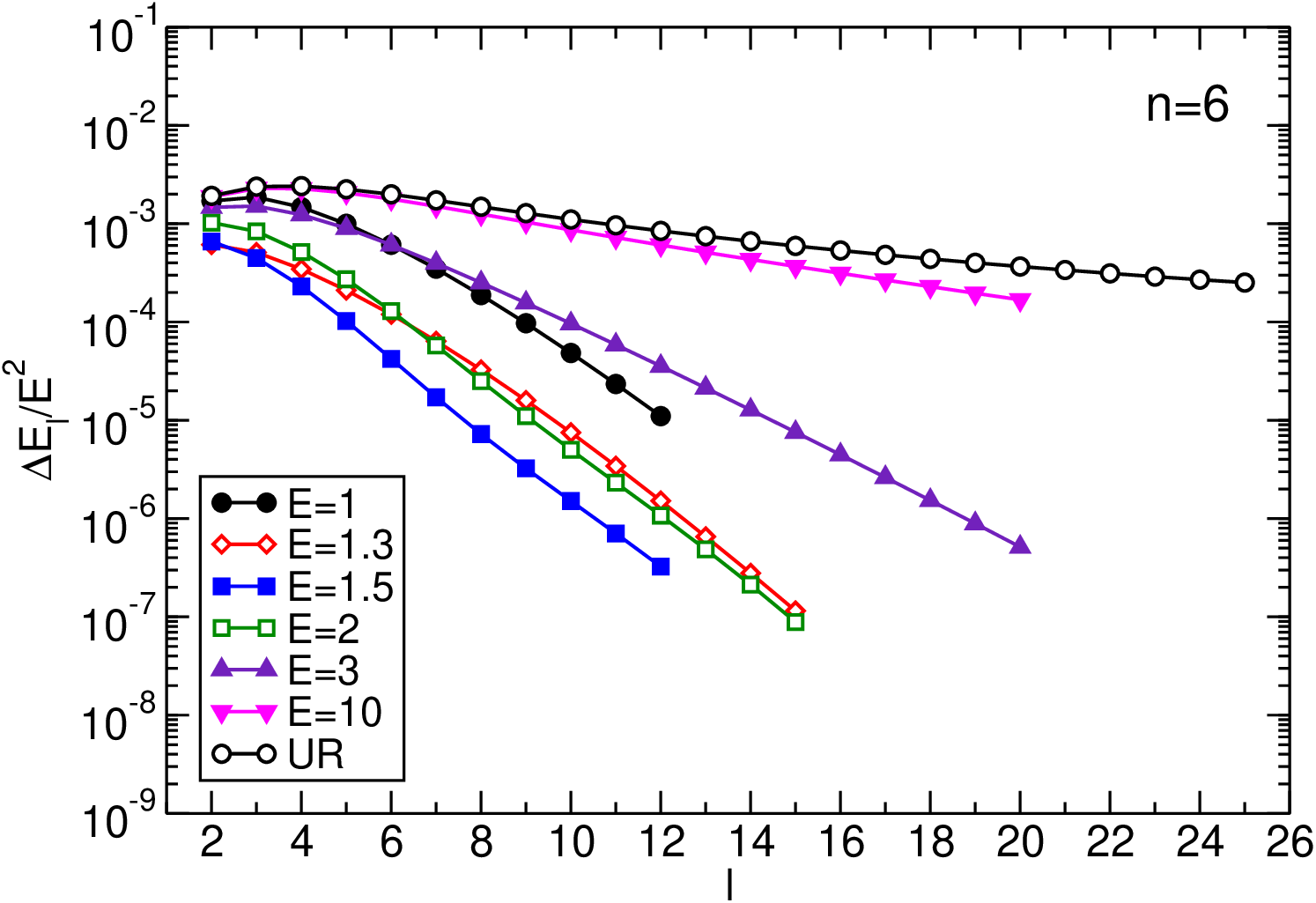}
&\includegraphics[scale=0.23,clip=true]{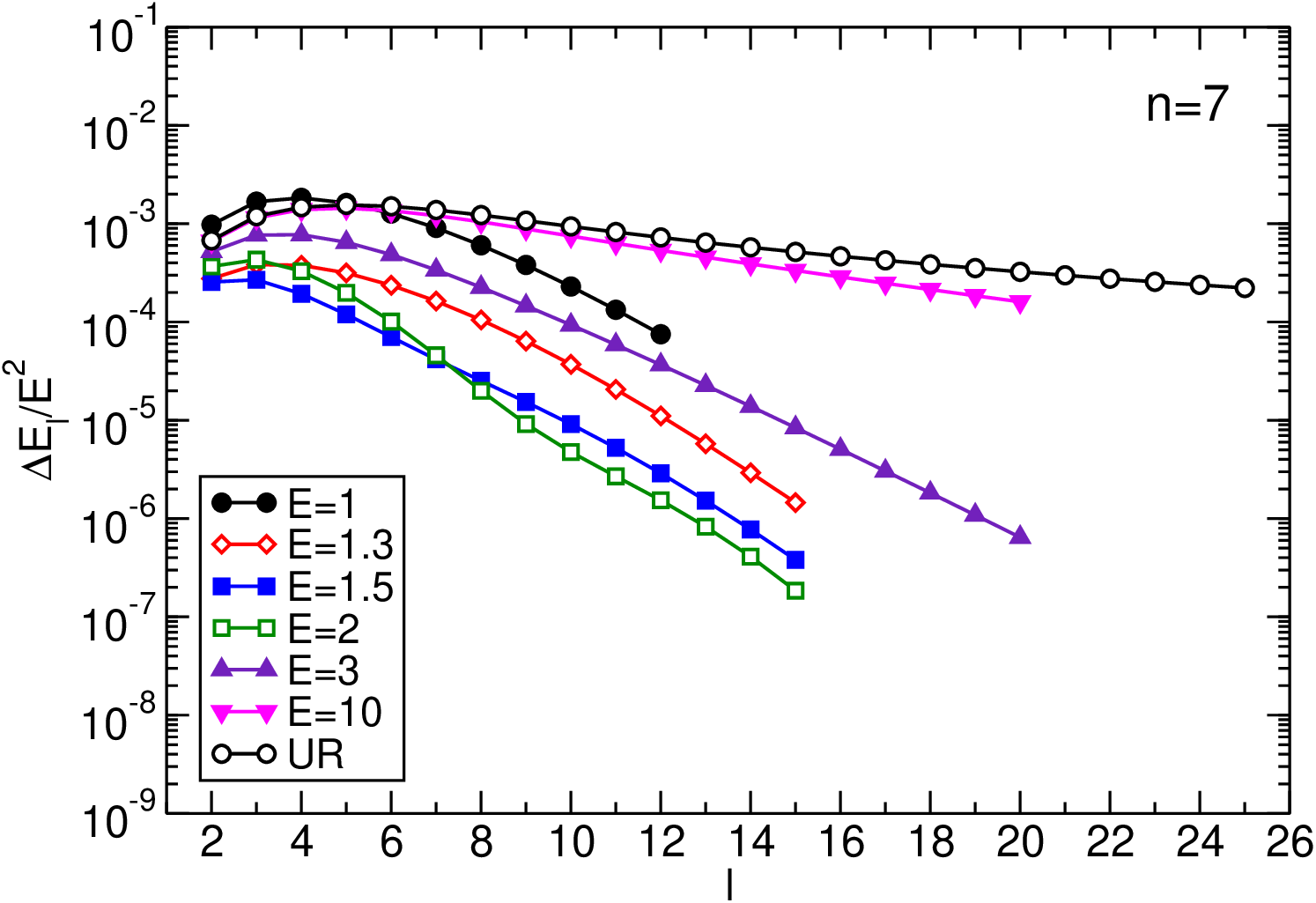}\\
\end{tabular}
\begin{tabular}{cc}
\includegraphics[scale=0.23,clip=true]{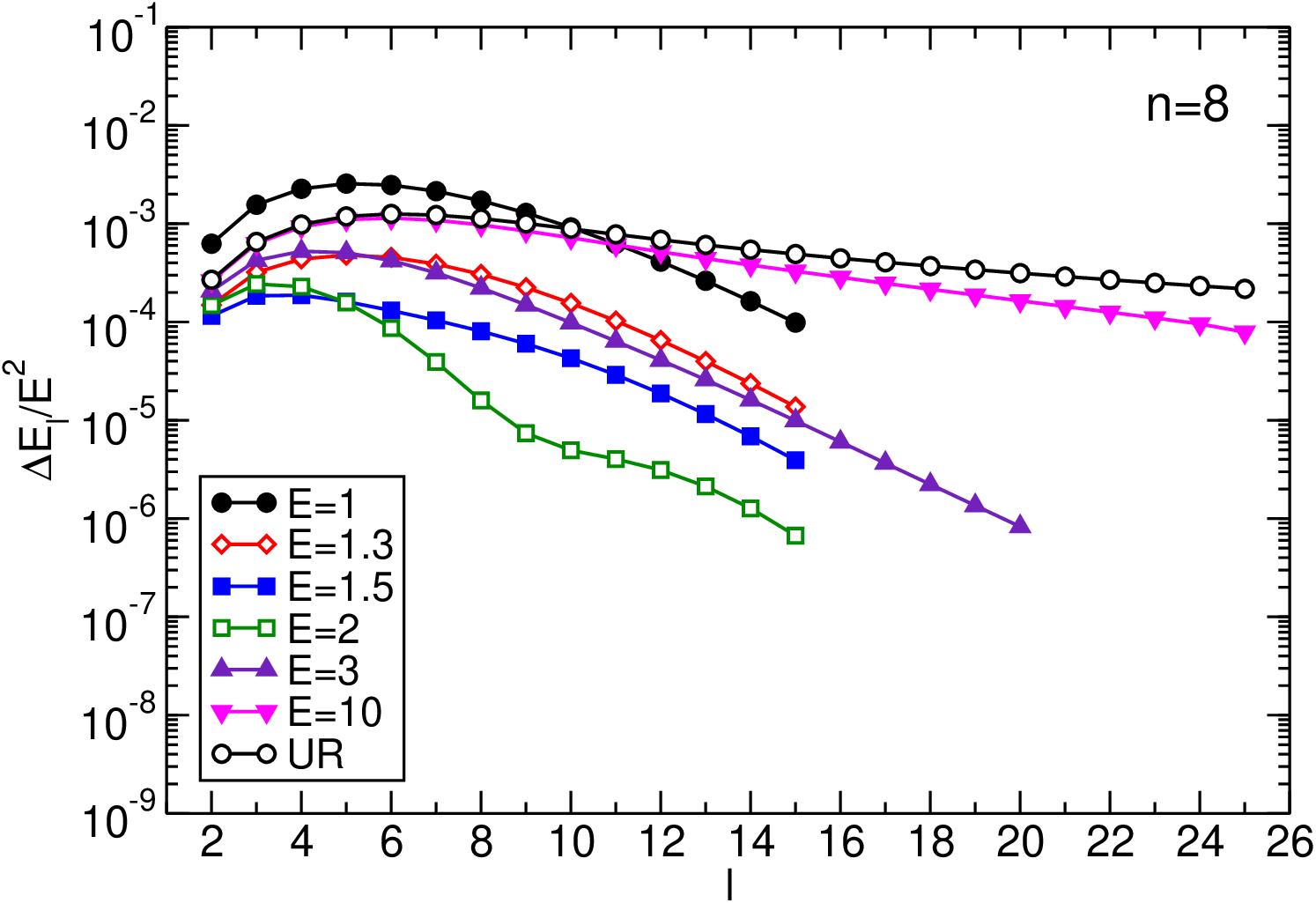}
&\includegraphics[scale=0.23,clip=true]{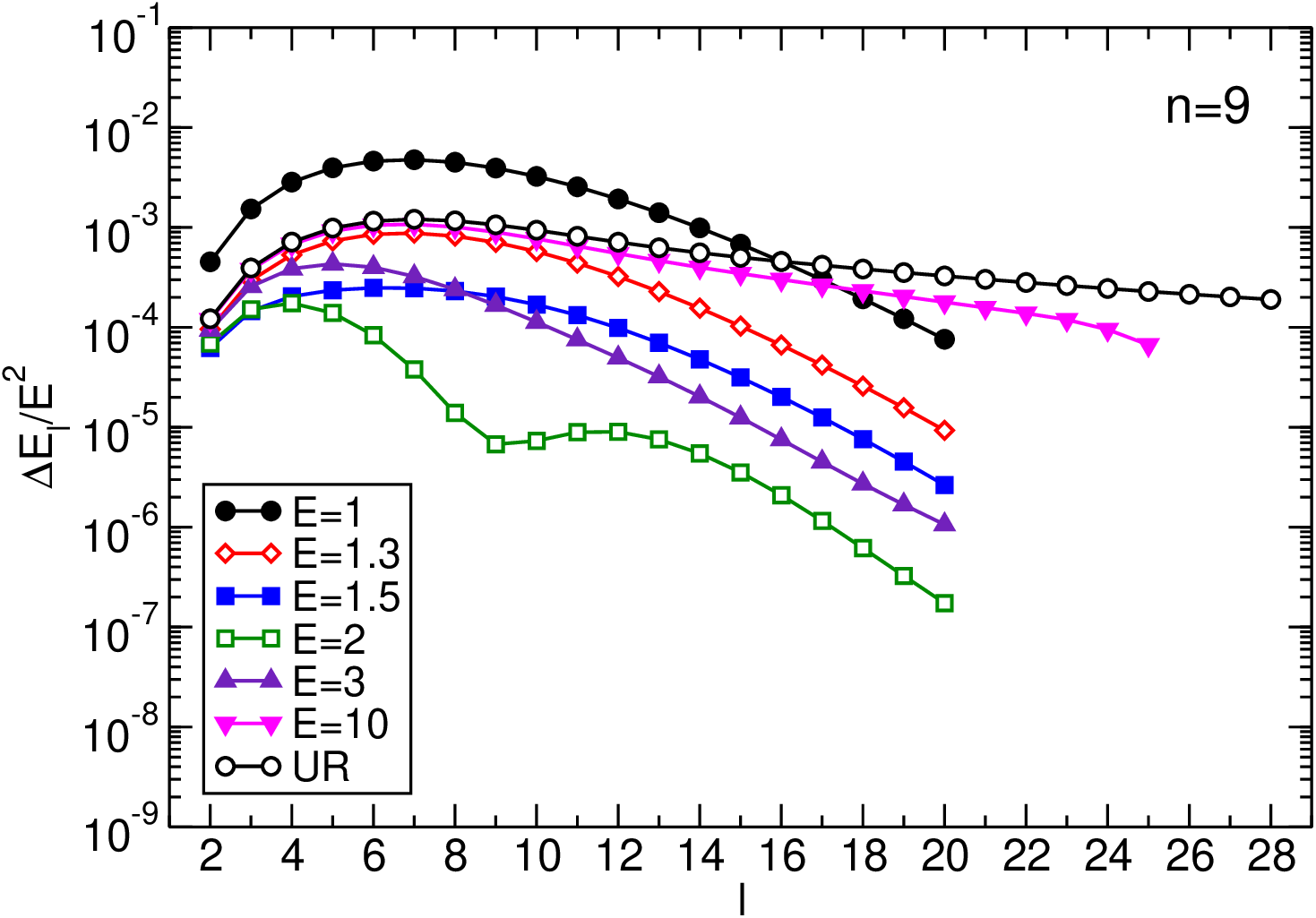}\\
\end{tabular}
\caption{\label{multipoles} Multipolar distribution for $n=3\,,\dots\,,9$.}
\end{center}
\end{figure*}

Figure \ref{n3spectra} shows representative energy spectra for $n=3$ and $n=6$
at different values of the particle energy.  In the ultrarelativistic limit,
as pointed out analytically in Ref.~\cite{Cardoso:2002pa} using Weinberg's
``zero-frequency limit'' approximation and confirmed numerically in Paper I,
at low frequencies the spectra grow like $\omega^{n-2}$, then they fall off
exponentially beyond a cutoff frequency $\omega_c$ corresponding to the
fundamental quasinormal mode frequency for the multipole in question
(cf. Figure 1 in Paper I). This can be understood in terms of
gravitational-wave scattering from the potential barrier surrounding the
BH. The quantity $\omega^2$ plays the role of the energy in the
Schr\"odinger-like equation (\ref{inhom}), so $\omega^2$ is equal to the
maximum of the scalar potential $V_0$ at first order in the WKB
approximation. Therefore, only the radiation with energy smaller than the peak
of the potential is backscattered to infinity; radiation with larger frequency
is exponentially suppressed. This interpretation explains the salient features
of Figure \ref{n3spectra}, and it is useful even in the context of
comparable-mass, ultrarelativistic BH collisions \cite{Berti:2010ce}.

A curious new feature of the energy spectra for $n\geq 5$ is the appearance of
a double peak for large multipole number and intermediate particle
energies. We have no quantitative explanation for these double peaks, but we
suspect that they may be somehow related to the appearance of multiple peaks
in the scalar potential for low values of $l$ (cf. Figure \ref{Vkodama}).

For a given particle energy, higher multipoles contribute more as $n$ grows.
This is even more evident when we look at the $\omega$-integrated multipolar
components of the energy spectra of Figure \ref{multipoles}. Starting from
$n=6$, in general the dominant multipole is no longer the
quadrupole. 

The total emitted energy is obtained by numerically integrating the spectra
over $\omega$ and then by summing the individual multipolar components $\Delta
E_l$, which are shown in Figure \ref{multipoles}. In principle, to compute
the total energy we need to carry out a sum of all values of $l$ up to $l\to
\infty$. It is of course impossible to compute multipolar contributions
$\Delta E_l$ for all values of $l$, so we computed a large enough number of
multipoles for any given dimensionality $n$ and particle energy $E$. In
practice, for large $l$ we fit the integrated $\Delta E_l$ with a power law of
the form
\be\label{Efit}
\Delta E_l=a_{n+2} l^{-b_{n+2}}\,,
\ee
where the coefficients $(a_{n+2}, b_{n+2})$ are obtained by fitting
(typically) the last five data points of each multipolar distribution in
Figure \ref{multipoles}. For each $n$ and $E$, the number of multipoles shown
in the figure was chosen to minimize the dependence of these fits (and of the
resulting extrapolation) on the specific values of $l$ chosen for the
fit. This extrapolation introduces larger uncertainties when $E$ and/or $n$
get large. Our final results are summarized in Table \ref{tab:energy} and in
the left panel of Figure \ref{fig:energymom}.

\begin{table}[hbt]
\centering \caption{Total energy radiated computed using the extrapolation of
  Eq.~(\ref{Efit}). For $E=2$ we actually extended the calculation up to
  $n=11$, and we found that a local minimum in the radiation occurs at $n=10$:
  the corresponding entries in this table would be $0.326$ ($n=10$) and
  $0.575$ ($n=11$). For $E=1$ and $n=11$ the corresponding entry in this table
  would be $123$, so the assumptions underlying our calculation are
  invalid (see text).} \vskip 12pt
\begin{tabular}{@{}cccccccc@{}}
\hline \hline
\multicolumn{8}{c}{$10^2\times (M_{\rm BH} E^{\rm rad})/(m_0^2E^2)$}\\ \hline
$n$ &$E=1$ &$E=1.3$ &$E=1.5$ &$E=2$ &$E=3$ &$E=10$ &UR\\
\hline \hline
2 & 1.04  & 2.19 & 3.52 & 6.49 & 11.9  & 23.5 & 26.2\\
3 & 1.65  & 1.87 & 2.75 & 5.36 & 9.47  & 18.2 & 24.9\\
4 & 2.02  & 1.32 & 1.75 & 3.46 & 6.48  & 13.9 & 19.8\\
5 & 2.31  & 0.905& 1.00 & 1.99 & 4.11  & 10.4 & 16.5\\
6 & 2.92  & 0.760& 0.598& 1.14 & 2.67  & 8.20 & 13.0\\
7 & 4.54  & 0.906& 0.457& 0.684& 1.88  & 6.99 & 11.4\\
8 & 8.27  & 1.52 & 0.545& 0.449& 1.44  & 5.98 & 10.9\\
9 & 17.7  & 3.16 & 1.00 & 0.330& 1.20  & 5.45 & 10.6\\
%
%2 & 1.041765680 & 3.518543890  & 11.87844990 & 23.49285700 & 26.2\\
%3 & 1.648843818 & 2.745993655  & 9.467819339 & 18.20852145 & 24.88304478\\
%4 & 2.024611079 & 1.745280431  & 6.479923975 & 13.91892116 & 19.78313783\\
%5 & 2.306519762 & 0.9973114806 & 4.113040595 & 10.36102771 & 16.48248417\\
%6 & 2.920013050 & 0.5977732652 & 2.673550270 & 8.197154580 & 12.98712369\\
%7 & 4.542304852 & 0.4570005514 & 1.875835900 & 6.986894003 & 11.36147391\\
%8 & 8.267924337 & 0.5445315633 & 1.437878623 & 5.982722477 & 10.88443226\\
%9 & 17.70254126 & 0.9959179753 & 1.197058570 & 5.449899469 & 10.63512154\\
%
\hline \hline
\end{tabular}
\label{tab:energy}
\end{table}

For $E=1$ (infall from rest) our results are well fitted by an expression of
the form
\be
10^2\times \frac{M_{\rm BH}}{m_0^2} E^{\rm rad}=
c_1 +c_2 \times c_3^{D}\,,
\label{fit}
\ee
where $c_1=1.865$, $c_2=8.037\times 10^{-4}$ and $c_3=2.457$.  Now, based
solely on the amount of emitted energy, one might expect the point-particle
approximation to break down when $E^{\rm rad}>m_0$. Based on the extrapolation
of Eq.~(\ref{fit}), this effectively constrains the mass ratio of the system
to values $m_0/M_{\rm BH}<1$ when $D>13$.  For smaller $D$ such a constraint
does not apply. This may help to explain some results in the literature. For
instance, consider the good agreement between numerical relativity simulations
of equal-mass BH collisions and the point-particle extrapolations to
equal-mass systems.  In $D=4$, early work \cite{Anninos:1993zj} and more
recent simulations (see e.g. \cite{Witek:2010xi}) found that the energy
radiated in full nonlinear simulations of head-on collisions with finite mass
ratio is well reproduced by simply taking the {\it linearized} result for a
particle falling into a BH and replacing $m_0\to \mu$, where $\mu$ is the
reduced mass of the system. This surprising agreement is at least consistent
with the observation that radiation is weak, and therefore nonlinear effects
are small.

However, as $D$ grows the amount of radiation emission also grows (a similar
effect was observed recently in other settings
\cite{Gal'tsov:2009zi,Gal'tsov:2010me}). The extrapolation of perturbative
results to finite mass ratio must eventually break down, for the following
reason. For two equal-mass BHs, Hawking's area theorem implies that the area
of the final BH must be equal to or larger than that of the initial BHs. This
implies the following bound on radiation emission \cite{Witek:2010xi}
\be
\frac{E^{\rm area}}{2M_{BH}}\leq 1-2^{-\frac{1}{D-2}}\,.
\ee
Now, this bound decreases with dimension and will eventually be a strong
restriction to the amount of radiation, violating the extrapolation of
Eq.~(\ref{fit}) to the equal-mass case. In fact, Hawking's area theorem is
more restrictive than Eq.~(\ref{fit}) for $D>13$ -- precisely when we know
that the point-particle approximation breaks down due to the large amount of
gravitational radiation.

Summarizing, our results are self-consistent, they are consistent with the area
theorem bound, and moreover they predict a qualitative change in the
equal-mass BH collision as $D$ increases: for $D\leq 13$ the amount of energy
lost as gravitational waves increases with $D$, but our calculation breaks
down at $D=13$, and presumably for higher dimensions the radiated energy may
start to decrease to conform with the area bound.

Ongoing simulations of head-on BH collisions from rest ($E=1$) in $D=5$
consider unequal-mass BHs with mass ratios $q=m_1/m_2$ in the range between
$1$ and $1/4$ \cite{witek2010}. Extrapolation of the numerical results to the
extreme mass ratio limit yields \cite{witek2010}
\be
E^{\rm rad}=0.0164\frac{m_0^2}{M_{\rm BH}}\,.
\ee
This prediction agrees within better than $1\%$ with the $n=3$, $E=1$
prediction listed in Table \ref{tab:energy}. This excellent agreement provides
a strong sanity check of the complex numerical relativity simulations, and a
useful example of the significance of point-particle calculations such as
those presented here.  A thorough analysis of the nonlinear simulations
(including more extensive comparisons with the point-particle limit) is in
preparation \cite{witek2010}.

%%%%%%%%%%%%%%%%%%%%%%%%%%%%%%%%%%%%%%%%%%%%%%%%%%%%%%%%%%%%%%%%%%%%%
\subsection{Linear momentum}
%%%%%%%%%%%%%%%%%%%%%%%%%%%%%%%%%%%%%%%%%%%%%%%%%%%%%%%%%%%%%%%%%%%%%

%
\begin{table}[hbt]
\centering \caption{Momentum radiated.} \vskip 12pt
\begin{tabular}{@{}cccccccc@{}}
\hline \hline
\multicolumn{8}{c}{$10^2\times (M_{\rm BH} P^{\rm rad})/(m_0^2E^2)$}\\ \hline
$n$ &$E=1.0$&$E=1.3$ &$E=1.5$ &$E=2$   &$E=3$ &$E=10$ &UR\\
\hline \hline
2 & 0.082    &0.22    &0.42     &1.1   &2.4   &5.9  &8.1 \\
3 & 0.26    &0.25    &0.43     &1.1   &2.6   &6.8  &9.3\\
4 & 0.51    &0.24    &0.32     &0.82   &2.1   &6.2  &8.3 \\
5 & 0.85    &0.25    &0.24     &0.59   &1.6   &5.4  &7.9 \\
6 & 1.4    &0.31    &0.20     &0.43   &1.3   &4.9  &7.3 \\
7 & 2.4    &0.47    &0.20     &0.31   &1.1   &4.5  &6.2 \\
8 & 4.7    &0.85    &0.28     &0.23   &0.92   &4.2  &5.6 \\
9 & 10     &1.8    &0.56     &0.18   &0.81   &4.1  &5.3 \\
\hline \hline
\end{tabular}
\label{tab:momentum}
\end{table}

In Table \ref{tab:momentum} and in the right panel of Figure
\ref{fig:energymom} we summarize the results for the linear momentum emitted
in gravitational waves. The pattern for momentum emission closely mimics that
of energy emission. If perturbative results can be extrapolated to finite mass
ratios (which is the case for lower spacetime dimensions, see
Ref.~\cite{witek2010}) one expects the following mass ratio dependence
\cite{Madalenathesis}:
\be
P=A_v (m_0E)\frac{q(1-q)}{(1+q)^5}\,,
\ee
where $q\equiv m_0E/M_{\rm BH}$. The quantity $A_v$ can be read off from Table
\ref{tab:momentum} in the small-$q$ limit.  From the momentum, one can get the
recoil velocity
\be
\frac{v_{\rm kick}}{c}=A_v \frac{q^2(1-q)}{(1+q)^5}\,.
\ee
This equation predicts a maximum kick velocity $v_{\rm kick}^{max}/c=0.0179
A_v$ for $q=(3+\sqrt{5})/2 \sim 0.382$. Numerical simulations of BH
collisions from rest in $D=5$ indicate that, in the point-particle limit
\cite{witek2010},
\be
\frac{v_{\rm kick}}{c}=0.24 \frac{m_0^2}{M_{\rm BH}}\,,
\ee
again in very good agreement with Table~\ref{tab:momentum}. This is a
nontrivial test of the simulations, because the emission of linear momentum
involves interference between different multipoles.

%%%%%%%%%%%%%%%%%%%%%%%%%%%%%%%%%%%%%%%%%%%%%%%%%%%
\section{\label{concl}Conclusions and outlook}
%%%%%%%%%%%%%%%%%%%%%%%%%%%%%%%%%%%%%%%%%%%%%%%%%%%

Our results for the energy spectrum, total energy and momentum radiated during
the head-on infall of a point particle into a higher-dimensional BH show an
interesting and complex structure.  The results indicate a beautiful
concordance with the area theorem and they suggest that the extrapolation of
perturbation theory to equal-mass collisions will yield wrong results for
dimensions $D \gtrsim 13$. This suggests that there should be a mechanism
suppressing the total amount of radiation in large spacetime dimensions. Full
nonlinear evolutions of the Einstein equations will probably be needed to
clarify the exact nature of this mechanism.  

A natural and interesting generalization of our results would be to study the
large-$D$ limit with either numerical or analytical techniques. Other obvious
generalizations include the study of infalls with finite impact parameters and
of rotating (Myers-Perry) black holes.

The present results should be relevant to the nascent field of numerical
relativity in higher-dimensional spacetimes. They can be used as a guide and
benchmark for future nonlinear simulations. Indeed, we will show in
forthcoming work how full numerical simulations of Einstein's equations are
remarkably consistent with the results reported here \cite{witek2010}.

%%%%%%%%%%%%%%%%%%%%%%%%%%%%%%%%%%%%%%%%%%%%%%%%%%%%%%%%%%%%%%%%%%%%%%%%%%%%%%
{\bf \em Acknowledgements.}
%%%%%%%%%%%%%%%%%%%%%%%%%%%%%%%%%%%%%%%%%%%%%%%%%%%%%%%%%%%%%%%%%%%%%%%%%%%%%%
We thank M. Cavagli\`a, L. Gualtieri, C. Herdeiro, A. Nerozzi, C. Ott,
U. Sperhake, N. Tufnel, H. Witek and M. Zilh\~ao for useful comments,
discussions and inspiration.  E.B. and B.K.'s research was supported by the
NSF under Grant No. PHY-0900735. V.C. acknowledges support from the
``Ci\^encia 2007'' program. This work was supported by the {\it DyBHo--256667}
ERC Starting Grant, NSF PHY-090003 and FCT - Portugal through PTDC projects
FIS/098025/2008, FIS/098032/2008, CTE-AST/098034/2008, and
CERN/FP/109290/2009. The authors thankfully acknowledge the computer
resources, technical expertise and assistance provided by the Barcelona
Supercomputing Centre---Centro Nacional de Supercomputaci\'on, as well as by
the DEISA Extreme Computing initiative and the Milipeia cluster in Coimbra.
%%%%%%%%%%%%%%%%%%%%%%%%%%%%%%%%%%%%%%%%%%%%%%%%%%%%%%%%%%%%%%%%%%%%%%%%%%%%%%%%

\appendix

\section{\label{app:source}Normalization coefficients and ultrarelativistic limit of the source
  term}

For the reader's convenience, here we list the normalized Gegenbauer
polynomials $\mathcal{S}^{nl}$ appearing in Eq.~(\ref{Gensource}) for the
relevant values of $n$ and $\theta=0$:
\begin{align}
\mathcal{S}^{2l}&=\frac{1}{2}\sqrt{\frac{2l+1}{\pi}}\,,\quad
\mathcal{S}^{3l}=\frac{l+1}{\pi\sqrt{2}}\,,\\
\mathcal{S}^{4l}&=\frac{1}{4\pi}\sqrt{(2l+3)\lambda_2}\,,\quad
\mathcal{S}^{5l}=\frac{1}{2\pi}\sqrt{\frac{(l+2)\lambda_3}{3\pi}}\,,\notag\\
\mathcal{S}^{6l}&=\frac{1}{8\pi}\sqrt{\frac{(2l+5)\lambda_4}{2\pi}}\,,\quad
\mathcal{S}^{7l}=\frac{1}{4\pi^2}\sqrt{\frac{(2l+6)\lambda_5}{15}}\,,\notag\\
\mathcal{S}^{8l}&=\frac{1}{16\pi^2}\sqrt{\frac{(2l+7)\lambda_6}{6}}\,,\quad
\mathcal{S}^{9l}=\frac{1}{4\pi^2}\sqrt{\frac{(l+4)\lambda_7}{105\pi}}\,,\notag\\
\mathcal{S}^{10l}&=\frac{1}{64\pi^2}\sqrt{\frac{(2l+9)\lambda_8}{6\pi}}\,,\quad
\mathcal{S}^{11l}=\frac{1}{24\pi^3}\sqrt{\frac{(2l+10)\lambda_9}{105}}\,,\notag
\end{align}
where $\lambda_k\equiv (l+k)(l+k-1)...(l+1)$. 

In the ultrarelativistic limit $E\rightarrow \infty$, the definition of the
tortoise coordinate implies that $t(r)=-r_*(r)$. In units
$r_h=(2M)^{1/(n-1)}=1$, the source term (\ref{Gensource}) reduces to:
\begin{align}
S_l^{(2)}&=e^{-i\omega r_*}
\frac{8\sqrt{4l+2}}{i\omega r}
\frac{(r-1)\nu_2}{\lb \nu_2 r+3 \rb^2}\,,\\ 
S_l^{(3)}&=e^{-i\omega r_*}
\frac{24(l+1)}{i\omega r^2\sqrt{\pi r}}
\frac{\lb (r^4-r^2)\nu_3+2(1-r^2)\rb}
{\lb \nu_3r^2+6 \rb^2}\nn\,,\\ 
S_l^{(4)}&=e^{-i\omega r_*}
\frac{16\sqrt{\lambda_2(2l+3)}}{i\omega r^3\sqrt{2\pi}}
\frac{\lb (r^6-r^3)\nu_4+5(1-r^3)\rb}
{\lb \nu_4r^3+10 \rb^2}\nn\,,\\ 
S_l^{(5)}&=e^{-i\omega r_*}
\frac{20\sqrt{\lambda_3(2l+4)}}{i\omega \pi r^3\sqrt{3r}}
\frac{\lb (r^8-r^4)\nu_5+9(1-r^4)\rb}
{\lb \nu_5r^4+15 \rb^2}\nn\,,\\ 
S_l^{(6)}&=e^{-i\omega r_*}
\frac{6\sqrt{\lambda_4(2l+5)}}{i\omega \pi r^4}
\frac{\lb (r^{10}-r^5)\nu_6+14(1-r^5)\rb}
{\lb \nu_6r^5+21 \rb^2}\nn\,,\\ 
S_l^{(7)}&=e^{-i\omega r_*}
\frac{28\sqrt{\lambda_5(2l+6)}}{i\omega \pi r^4\sqrt{30\pi r}}
\frac{\lb (r^{12}-r^6)\nu_7+20(1-r^6)\rb}
{\lb \nu_7r^6 +28\rb^2}\nn\,,\\ 
S_l^{(8)}&=e^{-i\omega r_*}
\frac{4\sqrt{\lambda_6(2l+7)}}{i\omega \pi r^5\sqrt{3\pi}}
\frac{\lb (r^{14}-r^7)\nu_8+27(1-r^7)\rb}
{\lb \nu_8r^7+36 \rb^2}\nn\,,\\ 
S_l^{(9)}&=e^{-i\omega r_*}
\frac{18\sqrt{\lambda_7(2l+8)}}{i\omega \pi^2r^5\sqrt{105r}}
\frac{\lb (r^{16}-r^8)\nu_9+35(1-r^8)\rb}
{\lb \nu_9r^8+45 \rb^2}\nn\,,\\
S_l^{(10)}&=e^{-i\omega r_*}\frac{5\sqrt{\lambda_8(2l+9)}}{i\omega
  4\pi^2r^6\sqrt{3}}
\frac{\lb (r^{18}-r^9)\nu_{10}+44(1-r^9)\rb}{\lb
  \nu_{10}r^9+55 \rb^2}\,,\nn\\
S_l^{(11)}&=e^{-i\omega r_*}\frac{22\sqrt{\lambda_9(2l+10)}}{i\omega
  3\pi^2r^6\sqrt{210\pi r}}\nn\\
&\times\frac{\lb (r^{20}-r^{10})\nu_{11}+54(1-r^{10})\rb}
{\lb \nu_{11}r^{10}+66 \rb^2}\,,\nn
\end{align}
where $\nu_k=(l+k)(l-1)$. These expressions are consistent with those listed
(for even dimensions) in Paper I.

\end{document}